\documentclass{jetpl}
\usepackage {cite}
\twocolumn

\lat

\title{Coulomb repulsion of holes and competition
between $d_{x^2-y^2}$-wave and $s$-wave parings in cuprate
superconductors}

\rtitle{Coulomb repulsion of holes and competition between
$d_{x^2-y^2}$- and $s$-wave parings in cuprate superconductors}

\sodtitle{Coulomb repulsion of holes and competition between
$d_{x^2-y^2}$- and $s$-wave parings in cuprate superconductors}

\author{V.\,V.\,Val'kov$^{a}$,
D.\,M.\,Dzebisashvili$^{a,b}$, M.\,M.\,Korovushkin$^{a}$, A.\,F.
Barabanov$^{c}$}

\rauthor{V.\,V.\,Val'kov, D.\,M.\,Dzebisashvili,
M.\,M.\,Korovushkin, and A.\,F. Barabanov}

\address{$^{a}$Kirensky Institute of Physics, Federal Research Center KSC SB RAS, 660036 Krasnoyarsk,
Russia\\
$^{b}$Reshetnev Siberian State University of Science and
Technology, 660037 Krasnoyarsk,
Russia\\
$^{c}$Vereshchagin Institute for High Pressure Physics, 108840
Troitsk, Russia}

\abstract{The effect of the Coulomb repulsion of holes on the
Cooper instability in an ensemble of spin-polaron quasiparticles
has been analyzed, taking into account the peculiarities of the
crystallographic structure of the CuO$_2$ plane, which are
associated with the presence of two oxygen ions and one copper ion
in the unit cell, as well as the strong spin–fermion coupling. The
investigation of the possibility of implementation superconducting
phases with $d$-wave and $s$-wave pairing of the order parameter
symmetry has shown that in the entire doping region only the
$d$-wave pairing satisfies the self-consistency equations, while
there is no solution for the $s$-wave pairing. This result
completely corresponds to the experimental data on cuprate HTSC.
It has been demonstrated analytically that the intersite Coulomb
interaction does not affect the superconducting $d$-wave pairing,
because its Fourier transform $V_q$ does not appear in the kernel
of the corresponding integral equation. }

\begin{document}

\maketitle

\section{INTRODUCTION}\label{sec1}

Analysis of specific properties of the normal phase of
high-temperature superconductors (HTSCs) leads to the conclusion
[1] that the insulator state of these materials is of the
Mott-Hubbard type [2, 3]. Accordingly, it was proposed that weakly
doped HTSCs be described on the basis of the Hubbard model [3] in
the strong electron correlation (SEC) regime. In [1], the
subsystem of the spin moments of copper ions was considered in
accordance with the scenario of resonant valence bonds, while the
charge excitation ensemble formed as a result of doping was
interpreted as the Fermi subsystem exhibiting the Cooper
instability. The mechanism of formation of the superconducting
phase appearing in such an approach was of the electronic origin
and led to high values of superconducting transition temperature
$T_c$.

Another solution to the problem of superconducting pairing with
high $T_c$ was proposed in [4], where it was shown that in the
range of low hole concentrations, the fermion ensemble described
by the Hubbard model in the limiting SEC regime
($U\rightarrow\infty$) exhibits the Cooper instability in the
$s$-wave channel. The new scenario of superconducting pairing was
based on the kinematic interaction that is initiated in the
Hubbard fermion ensemble due to the quasi-Fermi anticommutation
relations between the Hubbard operators [5]. The kinematic
mechanism of Cooper instability was also of the electron origin
and ensured high superconducting transition temperatures. The
inclusion of the intersite Coulomb interaction between fermions in
the Shubin–Vonsovsky model [6, 7] leads to a decrease in the
superconducting transition temperature [8-10] and gives
temperatures matching the experimental data.

The single-orbital Hubbard model, which basically reflects the
role of the SEC in the properties of the ground state and makes it
possible to analyze new mechanisms of Cooper instability in an
ensemble of strongly correlated fermions, disregarded specific
features of the crystalline structure of HTSCs. As a result,
important properties of the Fourier transforms of the matrix
elements for the intersite Coulomb repulsion, which are inherent
in the actual HTSC structure, were lost. This gave rise to the
problem (see below) associated with the strong suppression of the
superconducting phase with the $d$-wave type of the order
parameter symmetry in the case when the Coulomb repulsion of
fermions located at the nearest crystal lattice sites is taken
into account.

The model reflecting the actual structure of the CuO$_2$ plane was
formulated in [11]. The model took into account the fact that one
copper ion and two oxygen ions are located in the unit cell on the
CuO$_2$ plane. The inclusion of on-site Coulomb correlations made
it possible to pass to the SEC regime and to correctly describe
the Mott-Hubbard ground state of the system in the case of one
hole per unit cell. The papers [12-14] should also be mentioned in
this connection, in which the models that took into account the
HTSC structure, but differing either in the number of electron
orbitals of copper and the type of filling of electron orbitals
for Cu$^{3+}$ ions [13] or in the structure of included
interactions were proposed.

In the so-called Emery model that is used most frequently [11], it
was shown that the emergence of an additional hole in the CuO$_2$
plane leads to the formation of the spin–singlet state of the hole
located on the copper ion and an additional hole moving in the
oxygen binding orbital [15]. This stimulated attempts at
constructing an effective one-band model for cuprate HTSCs
[16-19].

Presuming that an effective Hubbard model or its low-energy
versions in the SEC limit must ultimately appear, most papers
devoted to the HTSC problem were based on the $t-J$ model on a
simple square lattice. In such an approach, the same fermions
formed both the charge and the spin subsystems, and the exchange
and spin-fluctuation mechanisms initiated Cooper pairing in the
$d$-wave channel [20-22].

Therefore, it seemed that the origin of the effective attraction
between the Hubbard fermions had basically been revealed. However,
the problem associated with the intersite Coulomb repulsion of
holes at oxygen remained unsolved. As a matter of fact, the Cooper
instability in the Hubbard model [4], $t-J$ model [21, 22], or
$t-J^*$ model [23, 24] was suppressed when the intersite Coulomb
repulsion of charge carriers was taken into account. This effect
manifested itself most strongly in the $d$-wave channel so that
superconductivity was suppressed completely for $V_1\sim1-2$ eV.
As a result, the contributions associated with the electron-
phonon, spin-fluctuation, and charge-fluctuations contributions
[25, 26] had to be taken into account additionally to compensate
the strong repulsion associated with the intersite Coulomb
interaction of holes. It should be noted, however, that the
Coulomb interaction potential between holes in different cells was
chosen in [25, 26] equal to $V = 0.2$ eV, which is much lower than
the spin–fluctuation pairing potential $g_{sf} = 1.5$ eV caused by
the kinematic interaction; it is only for this reason that the
superconducting $d$-wave phase was preserved. Due to a stronger
kinematic mechanism [4], Cooper pairing was also observed at
comparatively high values of $V$.

As a result, the following problem obviously arose: the
superconducting $d$-wave phase required for explaining
experimental results was strongly suppressed by the Coulomb
repulsion of holes located at the nearest sites. Note that
argumentation associated with the screening of the Coulomb
interaction, which is sometimes used in this connection, appears
as unconvincing in the given case because the repulsion of holes
at the shortest distances was considered [27]. Low effectiveness
of screening in HTSCs was noted in [14] and was associated with
the low concentration of holes at oxygen ions.

The problem of neutralization for the Coulomb repulsion of holes
at oxygen has required the revision of the existing theories of
Cooper instability in HTSCs. It should be noted in this connection
that an analogous problem also existed in the theory of classical
superconductors. Its solution had become possible after it was
shown [28, 29] that the electron-phonon interaction in a certain
region of the momentum space initiated effective attraction
between fermions, which could compensate for the bare repulsion.

It was shown in our recent paper [30] that the solution for the
problem of stability of the superconducting $d$-wave phase in
cuprates is associated with the rejection of the Hubbard model as
well as its low-energy modifications and with the return to the
model taking into account the actual structure of the CuO$_2$
plane in HTSCs. The role of such a model is played by the
spin-fermion model (SFM) formulated at early stages of development
of the HTSC theory [31-36]. This model follows directly from the
Emery model [11] if we take into account the effects of covalent
mixing of copper and oxygen orbitals in perturbation theory with
allowance for the actual relations between the initial Hamiltonian
parameters. Specific features of the SFM are associated with the
following factors. First, the SFM takes into account the spatial
separation of the subsystems of copper and oxygen ions (homeopolar
states of copper describe one hole). Second, which is significant,
the presence in the unite cell of two oxygen ions with the $p_x$
and $p_y$ orbitals is taken into account.

It was demonstrated in [30] that the allowance for the
above-indicated features of the SFM leads to the stability of the
phase with the $d_{x^2-y^2}$-wave symmetry of the order parameter
towards the strong Coulomb repulsion of holes located at the
nearest oxygen ions. However, the following two problems remain
unsolved: (i) the manifestation of the Coulomb interaction of
holes at the same oxygen ion in the problem of Cooper instability
and (ii) the competition of the superconducting $d$-wave and
$s$-wave phases. This study is devoted to the solution of these
problems.

This article is organized as follows. In Section 2, the Emery
model for cuprate superconductors is formulated. In Section 3, the
spin-fermion model is described, which follows from the Emery
model in the SEC regime. Section 4 is devoted to the derivation of
the equations for the normal and anomalous Green functions. The
system of integral equations for the superconducting order
parameter components is given in Section 5. In Section 6, the
influence of the Coulomb interaction on the evolution of the
Cooper instability in a spin-polaron ensemble is analyzed. The
stability of the superconducting $d$-wave pairing towards the
Coulomb repulsion of holes at the same and adjacent oxygen ions is
demonstrated. The competition between the $d$-wave and $s$-wave
pairings is investigated based on the calculated concentration
dependences of the superconducting transition temperature. In the
concluding section, the results of this study are discussed. For
convenience of presentation of the results, cumbersome analytic
expressions are given in Appendix A and Appendix B.

\begin{figure}
\begin{center}
\includegraphics[width=0.35\textwidth]{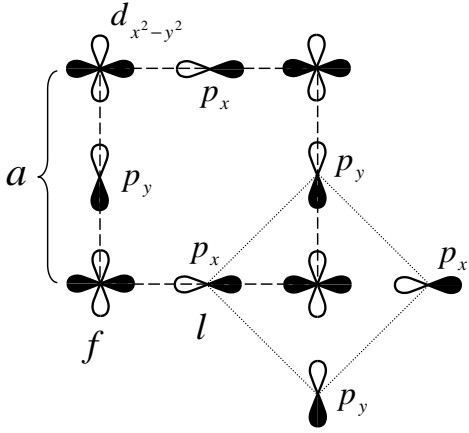}
\caption{Fig. 1. The $d_{x^2-y^2}$-orbitals of fermions on copper
ions and $p_x$ and $p_y$ orbitals of fermions at the oxygen ions
on the CuO$_2$ plane, which are taken into account in the Emery
model. The dashed lines bound the unit cell with parameter $a$.
The dotted lines connect four oxygen orbitals that are closest to
the copper orbital.} \label{fig-cell}
\end{center}
\end{figure}

\section{HAMILTONIAN OF THE EMERY MODEL}\label{sec2}

It is well known that the main features of the electronic
structure of the CuO$_2$ plane in HTSCs is correctly described by
the Emery model [11, 12, 14], in which the Hamiltonian in the
representation of the secondary quantization operators can be
written in the form
\begin{eqnarray}\label{Hpd}
&&\hat{H}=\hat{H}_0+\hat{U}_p+\hat{T}_{pd}+\hat{T}_{pp}+\hat{V}_{pp},\\
&&\hat{H}_0=\varepsilon_d \sum_{f}\hat n^d_{f}+U_d\sum_{f}\hat
n^d_{f\uparrow}\hat n^d_{f\downarrow} +\varepsilon_p \sum_{l}\hat
n^p_{l}\nonumber\\
&&\qquad+V_{pd}\sum_{f\delta}\hat n^d_{f} \hat n^p_{f+\delta},\nonumber\\
&&\hat{U}_p=U_p\sum_{l}\hat n^p_{l\uparrow}\hat
n^p_{l\downarrow},\qquad\hat{T}_{pp}=
\sum_{l\Delta\sigma}t_{pp}(\Delta)p^{\dagger}_{l\sigma}
p_{l+\Delta,\sigma},\nonumber\\
&&\hat{T}_{pd}=t_{pd}\sum_{f\delta\sigma}
\vartheta(\delta)(d^{\dagger}_{f\sigma}p_{f+\delta,\sigma}+\textrm{h.c.}),\nonumber\\
&&\hat{T}_{pp}=
\sum_{l\Delta\sigma}t_{pp}(\Delta)p^{\dagger}_{l\sigma}
p_{l+\Delta,\sigma},\nonumber\\
&&\hat{V}_{pp}=\sum_{ll'(l\ne l')}V_{pp}(l-l') \hat n^p_{l}\hat
n^p_{l'}.\nonumber
\end{eqnarray}
Here $d_{f\sigma}^ \dag (d_{f\sigma})$ and $p_{l\sigma}^ \dag
(p_{l\sigma})$ are the creation (annihilation) operators for the
$d$- and $p$-fermions, respectively, at copper $f$ and oxygen $l$
sites with spin projections $\sigma=+1/2,-1/2$. One of the four
vectors connecting the copper ion with the oxygen ions in the
CuO$_2$ plane is denoted by $\delta=\{\pm x/2,~\pm y/2 \}$, where
$x=(a,0)$ and $y=(0,a)$, $a$ being the unit cell parameter. Vector
$\delta$ connects the copper ion at site $f$ with the oxygen ions
in the position $l=f+\delta$ (Fig.~\ref{fig-cell}). The particle
number operators at copper and oxygen ions are defined as $\hat
n^d_f=\sum_{\sigma}\hat n^d_{f\sigma}=\sum_{\sigma}d_{f\sigma
}^\dag d_{f\sigma }$ è $\hat n^p_l=\sum_{\sigma}\hat
n^p_{l\sigma}=\sum_{\sigma}p_{l\sigma }^\dag p_{l\sigma }$. By
$\varepsilon_d$ and $\varepsilon_p$ we denote the bare on-site
energies of fermions on copper and oxygen ions, respectively.
Parameters $U_d$ and $U_p$ in the Hamiltonian indicate the Coulomb
repulsion energy for two particles with opposite spin projections
at a copper and an oxygen site, respectively; $V_{pd}$ is the
Coulomb repulsion energy for fermions at the copper and oxygen
ions, and $V_{pp}$ is the parameter of the Coulomb interaction of
fermions at oxygen ions. By $t_{pd}$ we denote the hopping
integral for a charge carrier from an oxygen ion to a copper ion.
Function $\vartheta(\delta)$ takes into account the effect of the
relation between the phases of copper and oxygen orbitals on the
hybridization processes. For the orbital profiles shown in
Fig.~\ref{fig-cell}, function $\vartheta(\delta)$ assumes the
following values upon the variation of $\delta$:
$\vartheta(\delta)=\mp 1$ for $\delta=\pm x/2$ or $\delta=\pm
y/2$~\cite{Zhang88}. By $t_{pp}(\Delta)=t\cdot\rho(\Delta)$, we
denote the fermion hopping integral between nearest oxygen
orbitals. Its sign is determined by function $\rho(\Delta)$, where
vector $\Delta$ connects the nearest oxygen ions. For the chosen
sequence of phases of oxygen orbitals, $\rho(\Delta)=1$ for
$\Delta=\pm(x+y)/2$ and $\rho(\Delta)=-1$ if $\Delta=\pm(x-y)/2$.

The Hamiltonian of the Emery model is a typical Hamiltonian in the
multiband theory of metals in the tight-binding approximation. It
belongs to the Hubbard type (the Emery model is often referred to
in the literature as the three-band or extended Hubbard model)
because it describes both intraatomic Coulomb correlations and
hopping between one-ion states of copper and oxygen. However, the
Emery model is more realistic as compared to the Hubbard model
because it takes into account the chemical composition of copper
oxides.

\section{SPIN-FERMION MODEL}
\label{sec:EMERY_model_SEC}

According to experimental data, in the undoped state with one hole
per unit cell, the system is in the state of a Mott-Hubbard
insulator~\cite{Mott79}. In the Emery model, this corresponds to
the SEC regime
\begin{equation}\label{cond_1}
\Delta_{pd},\,(U_d-\Delta_{pd}) \gg t_{pd}
>0.
\end{equation}
These inequalities require, on the one hand, that the Coulomb
correlations at the copper ion be taken into account correctly; on
the other hand, these inequalities make it possible to carry out
the reduction of the Hamiltonian in the Emery model and to obtain
the
SFM~\cite{Barabanov88,Prelovsek88,Zaanen88,Stechel88,Emery88,Matsukawa89}:
\begin{eqnarray}\label{Hamiltonianspinfermion}
&&\hat{H}_{\textrm{sp-f}}=\hat{H}_{\textrm{h}}+\hat{U}_{p}+\hat{V}_{pp}+\hat{J}+\hat{I},
\end{eqnarray}
where
\begin{eqnarray} &&\hat{H}_{\textrm{h}}=\sum_{k\alpha}\Bigl(\xi_{k_x}
a_{k\alpha}^{\dagger}a_{k\alpha}+ \xi_{k_y}
b_{k\alpha}^{\dagger}b_{k\alpha}\nonumber\\
&&\qquad+t_{k}( a_{k\alpha}^{\dagger}b_{k\alpha}+
b_{k\alpha}^{\dagger}a_{k\alpha})\Bigr),\label{def_Hh}\\
&&\hat{U}_{p}=\frac{U_p}{N}\sum_{1,2,3,4}
[a^{\dag}_{1\uparrow}a^{\dag}_{2\downarrow}a_{3\downarrow}a_{4\uparrow}+
(a\to b)
]~\delta_{1+2-3-4},\label{def_Up}\\
&&\hat{V}_{pp}=\frac{4V_1}{N}\sum_{1,2,3,4\atop{\alpha\beta}}\phi_{3-2}~
a^{\dag}_{1\alpha}b^{\dag}_{2\beta}b_{3\beta}a_{4\alpha}~\delta_{1+2-3-4}
,\label{def_V1}\\ \label{def_J}
&&\hat{J}=\frac{J}{N}\sum_{fkq\alpha\beta}e^{if(q-k)}
u_{k\alpha}^{\dag}(\textbf{S}_f\boldsymbol{\sigma}_{\alpha\beta})u_{q\beta},
\\
&&\hat{I}=\frac{I}{2}\sum_{
f\delta}\textbf{S}_f\textbf{S}_{f+2\delta}.
\end{eqnarray}
The relation between the operators of the oxygen subsystem in the
initial Emery model and the secondary quantization operators in
the SFM in the momentum representation is established by the
relations
\begin{eqnarray}\label{FT}
p_{f\pm\frac{x}{2},\sigma}=\frac{1}{\sqrt
N}\sum_{k}e^{ik(f\pm\frac{x}{2})}
a_{k\sigma},\nonumber\\
p_{f\pm\frac{y}{2},\sigma}=\frac{1}{\sqrt
N}\sum_{k}e^{ik(f\pm\frac{y}{2})} b_{k\sigma}.
\end{eqnarray}
Operators $a_{k\sigma}$ and $b_{k\sigma}$ correspond to the
annihilation of a hole with momentum $k$ and spin projection
$\sigma$, respectively, in the $x$- and $y$-sublattices of the
oxygen ions.

In the expression for the Hamiltonian $\hat{H}_{\textrm{h}}$, we
have introduced the functions
\begin{eqnarray}\label{xik}
\xi_{k_{x(y)}}=\varepsilon_p+2V_{pd}+\tau(1-\cos k_{x(y)})-\mu,
\end{eqnarray}
where $\mu$ is the chemical potential and parameter
$\tau=t^2_{pd}/\Delta_{pd}$. The function
\begin{eqnarray}\label{tk}
& t_{k}=(2\tau-4t)~\psi_k,\\ \label{psik} & \psi_k=s_{k,x}
s_{k,y},\qquad s_{k,x(y)}=\sin\frac{k_{x(y)}}{2}
\end{eqnarray}
describes the hybridization processes in the second order of
perturbation theory (parameter $\tau$) as well as direct hopping
of holes between the oxygen ions (parameter $t$). The dependence
for the sign of the hopping integrals on the direction of vector
$\Delta$ leads to the emergence of function $\psi_k$ in expression
(\ref{tk}).

For brevity, the momenta over which the summation is performed are
denoted by $1,\ldots,4$. The delta function $\delta_{1+2-3-4}$ in
the above expressions takes into account the momentum conservation
law. For operator $\hat{V}_{pp}$ of the intersite Coulomb
repulsion, we take into account the interactions only between the
nearest oxygen ions. The intensity of these interactions is
characterized by parameter $V_1$. Function $\phi_k$ in
$\hat{V}_{pp}$ is defined as
\begin{eqnarray}\label{phik}
\phi_k=\cos(k_x/2)\cos(k_y/2).
\end{eqnarray}
Operator $\hat J$ describes in the $k$ representation both
spincorrelated hopping of holes between oxygen ions and the
exchange interaction of a hole at the oxygen ion with the spins at
the nearest copper ions. Parameter $J$ of this interaction is
defined as $J=4t^2_{pd}/\Delta_{pd}$. In operator $\hat J$,
$\textbf{S}_f$ denotes the vector operator of the spin localized
at site $f$, while vector operator $\boldsymbol{\sigma}$ is
composed of the Pauli matrices: $\boldsymbol{\sigma} =
(\sigma^x,\sigma^y,\sigma^z)$. For brevity of notation, we have
introduced in expression (\ref{def_J}) the operator
\begin{eqnarray}\label{def_u}
\quad u_{k\beta}=s_{k,x}a_{k\beta}+ s_{k,y}b_{k\beta}.
\end{eqnarray}

The last term $\hat I$ in the Hamiltonian
(\ref{Hamiltonianspinfermion}) appears in the fourth order of
perturbation theory and describes the exchange interaction of
spins localized at copper ions.

The Hamiltonian in the SFM in the momentum representation was
considered earlier in~\cite{Starykh95}, where the spectrum of the
Fermi quasiparticles in Sr$_2$CuO$_2$Cl$_2$ was analyzed in the
self-consistent Born approximation. However, the Coulomb
interaction operators $\hat V_{pd}$, $\hat U_{p}$ and $\hat
V_{pp}$ were not taken into account.

When deriving expression (\ref{Hamiltonianspinfermion}) for the
Hamiltonian in the SFM, we assumed that the Coulomb repulsion
parameter for holes at copper ions was $U_d=\infty$. In further
analysis of the conditions for the evolution of the Cooper
instability in the SFM, we will use the wellestablished values of
parameters for the Emery model~\cite{Hybertsen89,Ogata08}:
$t_{pd}=1.3,\, \Delta_{pd}=3.6,\,U_p=4.0,\, V_{pd}=1.2$ (in eV).
For the hole hopping integral at oxygen, we will use the value of
$t=0.12$~eV \cite{Dzebisashvili13}, and the exchange interaction
constant between the spins of the copper ions is chosen to be
$I=0.136$ eV, which is in conformity with the available
experimental data on cuprate superconductors. The parameter of the
intersite Coulomb repulsion of holes is chosen as
$V_1=1\div2$~eV~\cite{Fischer11}.

\section{EQUATIONS FOR THE GREEN FUNCTIONS}\label{sec:equations}

An important feature of the SFM is that the exchange coupling
between localized spins and the spins of holes turns out to be
strong: $J=1.88$~eV $\gg\tau\approx 0.47$~eV. This means that when
describing the oxygen ion subsystem, the strong coupling between
holes at oxygen ions and the subsystems of spins at copper ions
must be taken into account exactly. For this purpose, it is
convenient to use the Zwanzig-Mori projection
method~\cite{Zwanzig61,Mori65,Roth68}. The method for calculating
the dispersion curves for spin-polaron excitations in the SFM,
which is based on this approach, was described in detail
in~\cite{Barabanov01} and was actively used in subsequent
studies~\cite{Dzebisashvili13,Val'kov14,Val'kov15}.

For taking into account the aforementioned strong spin-charge
coupling, it is necessary to introduce one more operator into the
basis set of operators (apart of operators $a_{k\alpha}$ and
$b_{k\alpha}$), viz.,
\begin{equation}\label{L_operator}
L_{k\alpha}=\frac1N\sum_{fq\beta} e^{if(q-k)}
(\textbf{S}_f\boldsymbol{\sigma}_{\alpha\beta})u_{q\beta}.
\end{equation}
For analyzing the conditions for Cooper instability, we must
supplement the above set of three operators with three extra
operators~\cite{Val'kov16,Val'kov15} ($\bar{\alpha}~=~-\alpha$)
\begin{eqnarray}\label{spin-fermion_basis_supercond}
a_{-k\bar{\alpha}}^{\dag},\quad b_{-k\bar{\alpha}}^{\dag},\quad
L_{-k\bar{\alpha}}^{\dag}.
\end{eqnarray}
The addition of these operators to the basis makes it possible to
analyze not only normal, but also anomalous thermodynamic means
using a unified approach.

The exact equations of motion for the first three basis operators
have the form
\begin{eqnarray}\label{Eq_a}
i\frac{da_{k\uparrow}}{dt}=\xi_{k_x}a_{k\uparrow}+t_{k}b_{k\uparrow}
+Js_{k,x}L_{k\uparrow}\nonumber\\
\qquad+\frac{U_p}{N}\sum_{1,2,3}a^{\dag}_{1\downarrow}a_{2\downarrow}a_{3\uparrow}\delta_{k+1-2-3}\nonumber\\
\qquad+\frac{4V_1}{N}\sum_{1,2,\alpha}
\phi_{k-2}~b^{\dag}_{1\alpha}b_{1-2+k,\alpha}a_{2\uparrow},
\end{eqnarray}
\begin{eqnarray}
i\frac{db_{k\uparrow}}{dt}=\xi_{k_y}b_{k\uparrow}+t_{k}a_{k\uparrow}
+Js_{k,y}L_{k\uparrow}\nonumber\\
\qquad+\frac{U_p}{N}\sum_{1,2,3}b^{\dag}_{1\downarrow}b_{2\downarrow}b_{3\uparrow}\delta_{k+1-2-3}\nonumber\\
+\frac{4V_1}{N}\sum_{1,2,\alpha}
~\phi_{1-2}~a^{\dag}_{1\alpha}a_{2\alpha}b_{1-2+k,\uparrow},\label{Eq_b}
\end{eqnarray}
\begin{eqnarray}
i\frac{dL_{k\uparrow}}{dt}=\sum_{q\beta}
(\textbf{S}_{k-q}\boldsymbol{\sigma}_{\uparrow\beta})
\left[\Bigl(\xi_{q_x} s_{q,x}+t_qs_{q,y}\Bigr)a_{q\beta}
\right.\nonumber\\ \left.+\Bigl(\xi_{q_y}
s_{q,y}+t_qs_{q,x}\Bigr)b_{q\beta}\right]
\nonumber\\
+\frac{U_p}{N}\sum_{1,2,3,4}\delta_{1-2+3-4}
\left[s_{1,x}(\textbf{S}_{k-1}\boldsymbol{\sigma}_{\downarrow\alpha})
a^{\dag}_{3\alpha}a_{4,\downarrow}a_{2\uparrow}\right.\nonumber\\
\left.
+s_{1,y}(\textbf{S}_{k-1}\boldsymbol{\sigma}_{\downarrow\alpha})
b^{\dag}_{3\alpha}b_{4,\downarrow}b_{2\uparrow}\right]\nonumber
\end{eqnarray}
\begin{eqnarray}\label{Eq_L}
+\frac{4V_1}{N}\sum_{1,2,3\atop{\alpha\beta}}\phi_{1-2}
\left[s_{1,x}(\textbf{S}_{k-1}\boldsymbol{\sigma}_{\uparrow\alpha})
b^{\dag}_{3\beta}b_{1-2+3,\beta}a_{2\alpha}\right.\nonumber\\
\left.
+s_{3,y}(\textbf{S}_{k-3}\boldsymbol{\sigma}_{\uparrow\alpha})
a^{\dag}_{1\beta}a_{2\beta}b_{1-2+3,\alpha}\right]\nonumber\\
+J\sum_{qp\alpha\beta}s^2_{p}
(\textbf{S}_{k-p}\boldsymbol{\sigma}_{\uparrow\alpha})
(\textbf{S}_{p-q}\boldsymbol{\sigma}_{\alpha\beta})u_{q\beta}
\nonumber\\
+\frac{iJ}{N}\sum_{pqq_1\atop{\alpha\beta\nu}}
(\boldsymbol{\sigma}_{\uparrow\nu}\times\boldsymbol{\sigma}_{\alpha\beta})
\textbf{S}_{k+p-q-q_1}u^{\dag}_{p\alpha}u_{q\beta}u_{q_1\nu}\nonumber\\
-4iI\sum_{qp\alpha}
\gamma_{1p}u_{q\alpha}\boldsymbol{\sigma}_{\uparrow\alpha}(\textbf{S}_{k-q+p}\times\textbf{S}_{-p}),
\end{eqnarray}
where $s^2_k=s^2_{k,x}+s^2_{k,y}$, and invariant $\gamma_{1p}$ of
the square lattice will be defined later (see relation
(\ref{gammas}) below). In addition, in equation of motion
(\ref{Eq_L}), we have introduced the Fourier transform of the spin
operator
$$\textbf{S}_{k}=\frac1N\sum_fe^{-ikf}\textbf{S}_{f}.$$

Within the projection method~\cite{Zwanzig61,Mori65}, the system
of the equations of motion for the Green functions can be written
as
\begin{eqnarray}\label{GF}
\omega~{\hat G}(k,\omega)=\hat K(k)+\hat D(k) \hat K^{-1}(k)~{\hat
G}(k,\omega),
\end{eqnarray}
where the matrix retarded Green function is defined by elements
${G}_{ij}(k,\omega)=\langle\langle
A_{ik}|A^{\dag}_{jk}\rangle\rangle_\omega$, and the elements of
the energy matrix $\hat D(k)$ and the normalization matrix $\hat
K(k)$ are defined as
\begin{eqnarray}\label{def_DK}
D_{ij}(k)=\langle\{[A_{ik},\hat H_{\textrm{sp-f}}],A^{\dag}_{jk}\}
\rangle,\nonumber\\
K_{ij}(k)=\langle\{A_{ik},A^{\dag}_{jk}\}\rangle.~~~
\end{eqnarray}
Operators $A_{ik}$ on the right-hand sides of these expressions
run through a set of six basis operators
$$\{a_{k\uparrow},b_{k\uparrow},L_{k\uparrow},
a^{\dag}_{-k\downarrow},b^{\dag}_{-k\downarrow},L^{\dag}_{-k\downarrow}
\},$$ and the angle brackets in relation (\ref{def_DK}) denote
thermodynamic mean.

Evaluating elements (\ref{def_DK}) and substituting them into
matrix equation (\ref{GF}), we obtain a closed system of equations
for the normal $G_{ij}$ and anomalous $F_{ij}$ Green functions
($j=1,2,3$)
\begin{eqnarray}\label{EqM_GF}
&&(\omega-\xi_{x})G_{1j} = \delta_{1j} + t_{k}G_{2j}+J_{x}G_{3j}
+\Delta_{1k}F_{1j}+\Delta_{2k}F_{2j},\nonumber\\
&&(\omega-\xi_{y})G_{2j} = \delta_{2j}+t_{k}G_{1j}+J_{y}G_{3j}
+\Delta_{3k}F_{1j}+\Delta_{4k}F_{1j},\nonumber\\
&&(\omega-\xi_{L})G_{3j} = \delta_{3j}K_{k}+
(J_{x}G_{1j}+J_{y}G_{2j})K_{k}
+\frac{\Delta_{5k}}{K_k}F_{3j},\nonumber\\
&&(\omega+\xi_{x})F_{1j} = \Delta_{1k}^*G_{1j}+\Delta_{3k}^*G_{2j}
-t_{k}F_{2j}+J_{x}F_{3j},\nonumber\\
&&(\omega+\xi_{y})F_{2j} =
\Delta_{2k}^*G_{1j}+\Delta_{4k}^*G_{2j}-
t_{k}F_{1j}+J_{y}F_{3j},\nonumber\\
&&(\omega+\xi_{L})F_{3j} =
\frac{\Delta^*_{5k}}{K_k}G_{3j}+(J_{x}F_{1j}+J_{y}F_{2j})K_{k}.
\end{eqnarray}
Here, we have introduced the following notation for the normal
Green functions:
\begin{eqnarray*}
&&G_{11}=\langle\langle
a_{k\uparrow}|a_{k\uparrow}^{\dag}\rangle\rangle_\omega,\qquad
G_{21}=\langle\langle
b_{k\uparrow}|a_{k\uparrow}^{\dag}\rangle\rangle_\omega,\nonumber\\
&&\qquad\qquad G_{31}=\langle\langle
L_{k\uparrow}|a_{k\uparrow}^{\dag}\rangle\rangle_\omega.
\end{eqnarray*}
Functions $G_{i2}$ and $G_{i3}$ ($i=1,2,3$) are defined
analogously, the only difference being that instead of
$a^{\dag}_{k\uparrow}$, we have operators $b^{\dag}_{k\uparrow}$
and $L^{\dag}_{k\uparrow}$, respectively. The anomalous Green
functions are defined as
\begin{eqnarray*}
&&F_{11}=\langle\langle
a_{-k\downarrow}^{\dag}|a_{k\uparrow}^{\dag}\rangle\rangle_\omega,\qquad
F_{21}=\langle\langle
b_{-k\downarrow}^{\dag}|a_{k\uparrow}^{\dag}\rangle\rangle_\omega,\nonumber\\
&&\qquad\qquad F_{31}=\langle\langle
L_{-k\downarrow}^{\dag}|a_{k\uparrow}^{\dag}\rangle\rangle_\omega.
\end{eqnarray*}
For $F_{i2}$ and $F_{i3}$ ($i=1,2,3$), we are using the same
notation for the second subscript.

When writing system of equations (\ref{EqM_GF}), we have used the
following functions:
\begin{eqnarray}\label{notations}
&&\xi_{x(y)}=\xi_{k_{x(y)}},\quad
J_{x(y)}=Js_{k,x(y)},\nonumber\\
&&\xi_L(k)=\tilde\varepsilon_p-\mu-2t+5\tau/2-J\nonumber\\
&&\quad\quad+[(\tau-2t)(-C_1\gamma_{1k}+C_2\gamma_{2k})\nonumber\\
&&\quad\quad+\tau(-C_1\gamma_{1k}+C_3\gamma_{3k})/2\nonumber\\
&&\quad\quad+JC_1(1+4\gamma_{1k})/4-IC_1(\gamma_{1k}+4)]K_{k}^{-1},
\end{eqnarray}
where $K_{k}
=\langle\{L_{k\uparrow},L^{\dag}_{k\uparrow}\}\rangle=3/4-C_1\gamma_{1k}$,
and $\gamma_{jk}$ denote invariants of the square lattice:
\begin{eqnarray}\label{gammas}
&&\gamma_{1k}=(\cos k_x+\cos k_y)/2,\nonumber\\
&&\gamma_{2k}=\cos k_x\,\cos k_y,\nonumber\\
&&\gamma_{3k}=(\cos2k_x+\cos 2k_y)/2.
\end{eqnarray}

For the components of the superconductor order parameter, which
are defined as
\begin{eqnarray}\label{def_Delta}
&&\Delta_{1k}=\langle\{[a_{k\uparrow},\hat H_{\textrm{sp-f}}],a_{-k\downarrow}\}\rangle,\nonumber\\
&&\Delta_{2k}=\langle\{[a_{k\uparrow},\hat H_{\textrm{sp-f}}],b_{-k\downarrow}\}\rangle,\nonumber\\
&&\Delta_{3k}=\langle\{[b_{k\uparrow},\hat H_{\textrm{sp-f}}],a_{-k\downarrow}\}\rangle,\nonumber\\
&&\Delta_{4k}=\langle\{[b_{k\uparrow},\hat H_{\textrm{sp-f}}],b_{-k\downarrow}\}\rangle,\nonumber\\
&&\Delta_{5k}=\langle\{[L_{k\uparrow},\hat
H_{\textrm{sp-f}}],L_{-k\downarrow}\}\rangle,
\end{eqnarray}
we obtain
\begin{eqnarray}
&&\Delta_{1k}=-\frac{U_p}{N}\sum_{q}\langle
a_{q\uparrow}a_{-q\downarrow}\rangle,
\nonumber \\
&&\Delta_{2k}=-\frac{4V_1}{N}\sum_{q}\phi_{k-q}\langle
a_{q\uparrow}b_{-q\downarrow}\rangle,
\nonumber \\
&&\Delta_{3k}=-\frac{4V_1}{N}\sum_{q}\phi_{k-q}\langle
b_{q\uparrow}a_{-q\downarrow}\rangle,\nonumber\\
&&\Delta_{4k}=-\frac{U_p}{N}\sum_{q}\langle
b_{q\uparrow}b_{-q\downarrow}\rangle, \nonumber
\end{eqnarray}
\begin{eqnarray}\label{Eq_Delta}
&&\Delta_{5k}=\frac {1}{N}\sum_{q}\biggl\{I_{k-q}\bigl(\langle
L_{q\uparrow}L_{-q\downarrow}\rangle -C_1\langle
u_{q\uparrow}u_{-q\downarrow}\rangle\bigr)\nonumber\\
&&\quad+8IC_1\langle u_{q\uparrow}u_{-q\downarrow}\rangle\biggr\}+
\frac {J}{N}\sum_{q}\biggl\{-2\gamma_{1q}\langle
L_{q\uparrow}L_{-q\downarrow}\rangle \nonumber\\
&&\quad+\bigl(3/2-4C_1\gamma_{1k}\bigr)\langle
u_{q\uparrow}u_{-q\downarrow}\rangle\biggr\}\nonumber\\
&&\quad-\frac {U_p}{N}\sum_q\biggl\{(3/8-C_1/2\cos k_x)\langle
a_{q\uparrow}a_{-q\downarrow}\rangle\nonumber\\
&&\quad+(3/8-C_1/2\cos k_y)\langle
b_{q\uparrow}b_{-q\downarrow}\rangle\biggr\}\nonumber\\
&&\quad-\frac
{V_1}{N}\sum_q\biggl\{(3/4-2C_1\gamma_{1k}+C_2\gamma_{2k})\psi_q\nonumber\\
&&\quad+C_2\sin k_x\sin k_y\phi_q\biggr\}\bigl(\langle
a_{q\uparrow}b_{-q\downarrow}\rangle+\langle
b_{q\uparrow}a_{-q\downarrow}\rangle\bigr)\nonumber\\
&&\quad+\frac{2}{N}\sum_{q}\left(\xi(q_x)s_{q,x}+t_qs_{q,y}
\right)\langle a_{q\uparrow}L_{-q\downarrow}\rangle\nonumber\\
&&\quad+\frac{2}{N}\sum_{q}\left(\xi(q_y)s_{q,y}+t_qs_{q,x}
\right)\langle b_{q\uparrow}L_{-q\downarrow}\rangle,
\end{eqnarray}
where $I_k=4I\gamma_{1k}$, and the mean is given by
\begin{eqnarray}
\langle u_{q\uparrow}u_{-q\downarrow}\rangle&=&-s^2_{q,x}\langle
a_{q\uparrow}a_{-q\downarrow}\rangle-s^2_{q,y}\langle
b_{q\uparrow}b_{-q\downarrow}\rangle\nonumber\\
&&-\psi_q\bigl(\langle
a_{q\uparrow}b_{-q\downarrow}\rangle+\langle
b_{q\uparrow}a_{-q\downarrow}\rangle\bigr).
\end{eqnarray}

When deriving expressions (\ref{notations}) and (\ref{Eq_Delta}),
we took into account the fact that the subsystem of spins
localized at copper ions is in the state of a quantum spin liquid.
In this case, spin correlation functions $C_j=\langle
S_0S_{r_j}\rangle$ appearing in expressions (\ref{notations}) and
(\ref{Eq_Delta}), satisfy the relations
\begin{equation}\label{spin_correlators}
C_j=3\langle S^x_0S^x_{r_j}\rangle=3\langle
S^y_0S^y_{r_j}\rangle=3\langle S^z_0S^z_{r_j}\rangle,
\end{equation}
where $r_j$ is the coordinate of the copper ion in the $j$th.
coordination sphere. In this case, $\langle S^x_f\rangle=\langle
S^y_f\rangle=\langle S^z_f\rangle=0$.

When deriving the fifth equation in (\ref{Eq_Delta}) for the mean
values of the product of operators that cannot be reduced to the
basis operators, we have used the relation
\begin{eqnarray}\label{REL}
\langle \left( \textbf{S}_f\boldsymbol{\sigma}_{\downarrow\alpha}
c_{k\alpha} \right) \left(
\textbf{S}_g\boldsymbol{\sigma}_{\uparrow\beta} c_{p\beta}
\right)\rangle~~~~~~~~~~~~~~\nonumber\\
=2\langle\left(\textbf{S}_f\textbf{S}_g\right)
c_{k\uparrow}c_{p\downarrow} \rangle-\langle \left(
\textbf{S}_f\boldsymbol{\sigma}_{\downarrow\alpha} c_{p\alpha}
\right) \left( \textbf{S}_g\boldsymbol{\sigma}_{\uparrow\beta}
c_{k\beta} \right)\rangle,
\end{eqnarray}
where summation over indices $\alpha$ and $\beta$ is implied.
Relation (\ref{REL}) is valid in the SU(2) invariant phase and
makes it possible to express this mean in terms of the mean value
of the basis operators. The anomalous mean $\langle
L_{q\uparrow}L_{-q,\downarrow}\rangle$ playing the decisive role
in the realization of the $d$-wave superconductivity in the
ensemble of spin-polaron quasiparticles appears in the sum in the
equation for the order parameter component $\Delta_{5k}$ in the
system (\ref{Eq_Delta}) only when relation (\ref{REL}) is used.For
thermodynamic means containing the scalar product of the spin
operators, the uncoupling procedure was used. This explains, in
particular, the emergence of magnetic correlator $C_1$, which is
proportional to exchange integral $I$, in the first term on the
right-hand side of the expression for $\Delta_{5k}$.

The contributions to $\Delta_{5k}$ from the intersite Coulomb
interaction immediately after the evaluation of the commutators
have the form
\begin{eqnarray}\label{D3_V1}
-\frac{4V_1}{N}\sum_{1,2,3,4\atop{\alpha\beta}}
\phi_{1-2}s_{1x}s_{3y}\delta_{1-2+3-4}
\nonumber\\
\times\left[\langle(\textbf{S}_{k-1}\boldsymbol{\sigma}_{\uparrow\alpha}a_{2\alpha})
(\textbf{S}_{-k-3}\boldsymbol{\sigma}_{\downarrow\beta}b_{4\beta})\rangle
\right. \nonumber\\
\left.+\langle(\textbf{S}_{k-3}\boldsymbol{\sigma}_{\uparrow\alpha}b_{4\alpha})
(\textbf{S}_{-k-1}\boldsymbol{\sigma}_{\downarrow\beta}a_{2\beta})\rangle
\right].
\end{eqnarray}
Since the operators in the mean cannot be reduced to basis
operators even when relation (\ref{REL}) is used, the uncoupling
procedure is employed for the mean values in expression
(\ref{D3_V1}) taking into account the SU(2) invariance of the spin
subsystem. This procedure leads to the emergence of the term
proportional to $V_1$ in the fifth equation in (\ref{Eq_Delta}).

It should also be noted that since we are interested in the weak
doping regime, the contributions appearing in expressions
(\ref{notations}) and (\ref{Eq_Delta}) as a result of uncoupling
of the means and proportional to correlators of the
density-density type are not considered here.

Analysis of system of equations (\ref{EqM_GF}) in the normal phase
leads to the conclusion that the Fermi excitation spectrum in the
SFM is determined by the solutions to the dispersion equation
\begin{eqnarray}\label{det}
&&\mathrm{det}_{k}(\omega)=(\omega-\xi_{x})(\omega-\xi_{y})
(\omega-\xi_{L})-2J_{x}J_{y}t_{k}K_{k}\nonumber\\
&&-(\omega-\xi_{y})J_{x}^2K_{k} -(\omega-\xi_{x})J_{y}^2K_{k}
-(\omega-\xi_{L})t_{k}^2=0,~
\end{eqnarray}
and contains three branches: $\epsilon_{1k}$, $\epsilon_{2k}$ and
$\epsilon_{3k}$~\cite{Val'kov15}. Lower branch $\epsilon_{1k}$ is
characterized by a minimum near point ($\pi/2$, $\pi/2$) of the
Brillouin zone and is separated considerably from the two upper
branches $\epsilon_{2k}$ and $\epsilon_{3k}$. The lower branch
appears due to the strong spin– charge coupling that induces the
exchange interaction between holes and localized spins at the
nearest copper ions, as well as spin-correlated hopping. At low
doping levels, the dynamics of holes at oxygen ions is determined
predominantly by the lower branch $\epsilon_{1k}$.

\section{SYSTEM OF EQUATIONS FOR THE SUPERCONDUCTING ORDER PARAMETER COMPONENTS}
\label{sec:equations_Delta}

For analyzing the conditions for the Cooper instability, let us
express the required anomalous Green functions in terms of
parameters $\Delta^*_{lk}$. in the linear approximation. These
functions have the form
\begin{equation}\label{Fij}
F_{ij}(k,\omega)=\sum_{i,j=1}^3\sum_{l=1}^5\frac{S_{ij}^{(l)}(k,\omega)}{{\rm
Det}_k(\omega)}\Delta_{lk}^*.
\end{equation}
The Green functions required for analyzing the conditions for the
emergence of superconductivity are $F_{11}(k,\omega)$,
$F_{12}(k,\omega)$, $F_{21}(k,\omega)$, $F_{22}(k,\omega)$,
$F_{31}(k,\omega)$, $F_{32}(k,\omega)$ and $F_{33}(k,\omega)$.
Here,
$\textrm{Det}_k(\omega)=-\textrm{det}_k(\omega)\textrm{det}_k(-\omega)$,
and corresponding functions $S_{ij}^{(l)}(k,\omega)$ are given in
Appendix A.

Using the spectral theorem~\cite{Zubarev60}, we obtain the
following expressions for anomalous means and the closed system of
homogeneous integral equations for the superconducting order
parameter components ($l=1,\ldots,5$)
\begin{eqnarray}\label{Deltas_spectral_theorem}
&&\Delta_{1k}^*=-\frac{U_p}{N}\sum_{lq}M^{(l)}_{11}(q)
\Delta_{lq}^*,\\
&&\Delta_{2k}^*=-\frac{4V_1}{N}\sum_{lq}\phi_{k-q}M^{(l)}_{21}(q)
\Delta_{lq}^*,\nonumber\\
&&\Delta_{3k}^*=-\frac{4V_1}{N}\sum_{lq}\phi_{k-q}M^{(l)}_{12}(q)
\Delta_{lq}^*,\nonumber\\
&&\Delta_{4k}^*=-\frac{U_p}{N}\sum_{lq}M^{(l)}_{22}(q)
\Delta_{lq}^*,\nonumber\\
&&\Delta_{5k}^*=-\frac1N\sum_{lq}R^{(l)}_0(q)\Delta_{lq}^*
+\frac1N\sum_{lq}I_{k-q}R^{(l)}_{1a}(q)\Delta_{lq}^*\nonumber\\
&&\quad+\cos k_x\frac1N\sum_{lq}R^{(l)}_{1b}(q)\Delta_{lq}^*+\cos
k_y\frac1N\sum_{lq}R^{(l)}_{1c}(q)\Delta_{lq}^*\nonumber\\
&&\quad-
\gamma_{2k}\frac1N\sum_{lq}\psi_qR^{(l)}_2(q)\Delta_{lq}^*\nonumber\\
&&\quad-\sin k_x\sin
k_y\frac1N\sum_{lq}\phi_qR^{(l)}_2(q)\Delta_{lq}^*,\nonumber
\end{eqnarray}
where the following functions have been introduced:
\begin{eqnarray}
&&R^{(l)}_0(q)=\displaystyle\frac34V_1\psi_q
M^{(l)}_{ab}(q)+2J\gamma_{1q}M^{(l)}_{33}(q)\nonumber\\
&&\quad-(8IC_1+3J/2)M^{(l)}_{uu}(q)+\displaystyle\frac38U_p(M^{(l)}_{11}(q)+M^{(l)}_{22}(q))\nonumber\\
&&\quad-2\left(\xi(q_x)s_{q,x}+t_qs_{q,y} \right)M^{(l)}_{31}(q)\nonumber\\
&&\quad-2\left(\xi(q_y)s_{q,y}+t_qs_{q,x}
\right)M^{(l)}_{32}(q),\nonumber
\end{eqnarray}
\begin{eqnarray}
&&R^{(l)}_{1a}(q)=M^{(l)}_{33}(q)-C_1M^{(l)}_{uu}(q),\nonumber\\
&&R^{(l)}_{1b}(q)=C_1(V_1\psi_q
M^{(l)}_{ab}(q)-2JM^{(l)}_{uu}(q)+U_pM^{(l)}_{11}(q)),\nonumber\\
&&R^{(l)}_{1c}(q)=C_1(V_1\psi_q
M^{(l)}_{ab}(q)-2JM^{(l)}_{uu}(q)+U_pM^{(l)}_{22}(q)),\nonumber\\
&&R^{(l)}_{2}(q)=V_1C_2
M^{(l)}_{ab}(q),\\ \nonumber\\
&&M^{(l)}_{uu}(q)=-s_{qx}^2M^{(l)}_{11}(q)-s_{qy}^2M^{(l)}_{22}(q)-\psi_qM^{(l)}_{ab}(q),\nonumber\\
&&M^{(l)}_{ab}(q)=M^{(l)}_{21}(q)+M^{(l)}_{12}(q),\nonumber\\
&&M^{(l)}_{nm}(q)=\frac{S^{(l)}_{nm}(q,E_{1q})+S^{(l)}_{nm}(q,-E_{1q})}
{4E_{1q}(E_{1q}^2-E_{2q}^2)(E_{1q}^2-E_{3q}^2)}
\tanh\left(\frac{E_{1q}}{2T}\right).\nonumber
\end{eqnarray}
System of equations (\ref{Deltas_spectral_theorem}) will be used
below for determining the temperature of transition of an ensemble
of polarons to the superconducting state with preset type of order
parameter symmetry.

\section{COMPETITION OF $d$- AND $s$-WAVE
PAIRINGS OF SPIN POLARONS TAKING INTO ACCOUNT THE COULOMB
INTERACTIONS}\label{sec:V}

It can be seen from system (\ref{Deltas_spectral_theorem}) that
the kernels of integral equations are uncoupled; therefore, the
solution to this system can be sought in the form
\begin{eqnarray}\label{eqs_solution}
&&\Delta_{1k}=B_1,\nonumber\\
&&\Delta_{2k}=B_{1\phi}\phi_k + B_{1\psi}\psi_k,\nonumber\\
&&\Delta_{3k}=B_{2\phi}\phi_k + B_{2\psi}\psi_k,\nonumber\\
&&\Delta_{4k}=B_2,\nonumber\\
&&\Delta_{5k}=B_3+B_{cx}\cos k_x+B_{cy}\cos k_y\nonumber\\
&&\qquad+B_{ññ}\cos{k_x}\cos{k_y}+ B_{ss}\sin{k_x}\sin{k_y},
\end{eqnarray}
where eleven amplitudes $B_j$ ($j=1,1\phi,1\psi,...$) determine
the contribution of the corresponding basis functions to the
expansion of the order parameter components. Substituting these
expressions into Eqs. (\ref{Deltas_spectral_theorem}) and equating
the coefficients of the corresponding trigonometric functions, we
obtain the system of eleven algebraic equations for determining
amplitudes $B_j$. Actually, the situation is simplified because
the system splits into two independent subsystems. The first
subsystem defines three amplitudes ($B_{1\phi}$, $B_{2\phi}$ and
$B_{ss}$). Numerical calculations show that in the entire doping
range of interest, this system has no solutions and will not be
considered here.

The second subsystem of equations defines the remaining eight
amplitudes $B_j$, which can be conveniently written in the form of
a column vector
\begin{eqnarray}\label{vec_B}
B=(B_3,B_{cx},B_{cy},B_{cc},B_{1\psi},B_{2\psi},B_1,B_2)^T.
\end{eqnarray}
In matrix form, the system of eight equations can be written as
\begin{eqnarray}\label{sys_B}
B=\hat{W}B,
\end{eqnarray}
where components of eighth-order matrix $\hat W$ can be calculated
using the expressions
\begin{eqnarray}\label{def_W}
W_{ij}=\frac1N\sum_k w_{ij}(k),~~~(i,j=1,\dots,8),
\end{eqnarray}
and functions $w_{ij}(q)$ are given in Appendix B.

To determine the dependence of superconducting transition
temperature $T_c$ on doping level $x$ for different types of the
order parameter symmetry, we should solve Eq. (\ref{sys_B})
together with the equation for chemical potential $\mu$. In
deriving the equation for $\mu$, we should take into account the
fact that all order parameters $\Delta_{jk}\to0$ in the limit of
interest $T\to T_c$. As a result, we obtain the following equation
for determining the chemical potential:
\begin{eqnarray}\label{Eq_mu}
x=\frac{2}{N}\sum_q\frac{f(\epsilon_{1q})\left[Q_{3x}(q,\epsilon_{1q})+Q_{3y}(q,\epsilon_{1q})\right]}
{\left(\epsilon_{1q}-\epsilon_{2q}\right)\left(\epsilon_{1q}-\epsilon_{3q}\right)},
\end{eqnarray}
where $f(E)=(e^{E/T}+1)^{-1}$ is the Fermi-Dirac distribution
function.
\begin{figure}[t]
\begin{center}
\includegraphics[width=0.5\textwidth]{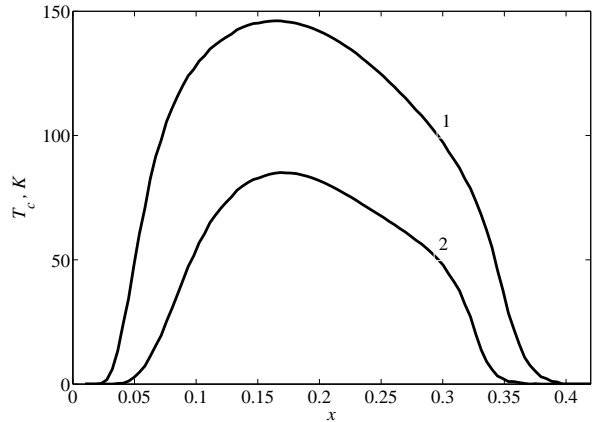}
\caption{Fig.~2. Doping dependences of the superconducting
transition temperature for the $d_{x^2-y^2}$-wave pairing,
obtained for the model parameters $J$=1.88, $\tau$ =0.47,
$t$=0.12, $I$=0.136. Curves 1 and 2 describe $T_c(x)$ for $U_p=0$
and $U_p=3$, respectively. The inclusion of intersite Coulomb
interaction $V_1$ does not influence on these dependencies. All
energy parameters are given in eV.}\label{fig-2}
\end{center}
\end{figure}

The results of numerical self-consistent solution of system of
equations (\ref{sys_B}) together with equation (\ref{Eq_mu}), for
the chemical potential are represented in Fig.~\ref{fig-2}. Solid
curve 1 shows the dependence of the critical temperature of
superconducting $d_{x^2-y^2}$-wave pairing on the doping level for
$U_p=0$ and $V_1=0$. This curve was obtained earlier
in~\cite{Val'kov15} and is in good agreement with experimental
data on the absolute value of $T_c$ and on the doping region in
which the Cooper instability evolves.

An important aspect of the approach developed here is that the
inclusion of Coulomb interaction $V_1$ of fermions located at the
nearest oxygen ions does not affect the $T_c(x)$ dependence for
the superconducting $d_{x^2-y^2}$-wave pairing: curve 1 in
Fig.~\ref{fig-2} remains unchanged \cite{Val'kov16}. The cause for
such a behavior can be found after analysis of the solutions to
system of integral equations (\ref{Deltas_spectral_theorem}). In
the doping interval in which the above type of pairing is realized
for $T\lesssim T_c$, the solutions to algebraic system
(\ref{sys_B}) for the amplitudes are such that only four
amplitudes $B_{cx}$, $B_{cy}$, $B_{1\psi}$ and $B_{2\psi}$ differ
from zero and $B_{cx}=-B_{cy}$, $B_{1\psi}=-B_{2\psi}$, and
$|B_{cx}|/|B_{1\psi}|\sim 10^3$. This means that the dependence of
the superconducting gap on $k$ is mainly due to the fifth
component $\Delta_{5k}$ of superconducting order parameter, which
in this case has the form
\begin{eqnarray}\label{Dcos}
\Delta^{(d)}_{5k}=B_{cx}\cdot(\cos k_x-\cos k_y).
\end{eqnarray}
Since for the $d$-wave pairing for $U_p=0$, amplitudes $B_{cx}$
and $B_{cy}$ in the equation for $\Delta_{5k}$ are determined not
by parameter $V_1$, but by the exchange coupling constant $I$
alone, we arrive at the conclusion that the Coulomb repulsion of
holes located at neighboring oxygen sites do not suppress the
superconducting phase with the $d_{x^2-y^2}$-wave order parameter
symmetry~\cite{Val'kov16}.

This means that in the case of the $d$-wave pairing and $U_p=0$,
we can obtain instead of system (\ref{sys_B}) a simpler equation
for the superconducting transition temperature
$T_c$~\cite{Val'kov15,VDB_JLTP_2015,VDB_JSNM_2016}. This equation
follows from the fifth equation of system
(\ref{Deltas_spectral_theorem}) and has the form
\begin{eqnarray}\label{Eq_Tc_d}
&&1=\frac{I}{N}\sum_q(\cos q_x-\cos q_y)^2\nonumber\\
&&\qquad\times\left(M_{33}^{(5)}(q,\epsilon_{1q})-C_1M_{uu}^{(5)}(q,\epsilon_{1q})\right).
\end{eqnarray}
This equation implies, in particular, that the exchange
interaction of spin moments of the copper ions, which is
transformed into effective attraction as a result of the strong
spin-charge coupling, is the mechanism of the Cooper instability.
The results of solution of Eq. (\ref{Eq_Tc_d}) and system
(\ref{sys_B}) for the $d$-wave pairing and $U_p=0$ obviously
coincide and correspond to solid curve 1 in Fig.~\ref{fig-2}.

In contrast to the intersite interaction, the allowance for the
Coulomb interaction $U_p$ of two holes at oxygen ion leads to the
suppression of the superconducting $d$-wave phase. However, as it
follows from comparison of the curve 2 ($U_p = 3$ eV) and the
curve 1 ($U_p=0$) in Fig.~\ref{fig-2}, this suppression is not
essential for the implementation of HTSC, since in the region of
optimal doping $x\simeq 0.16$ the critical temperature remains
high.

From the system of integral equations
(\ref{Deltas_spectral_theorem}) it follows that the solution
corresponding to the $s$-phase should have the form
\begin{eqnarray}\label{eqs_s}
&&\Delta^{(s)}_{1k}=\Delta^{(s)}_{4k}=B_1,\nonumber\\
&&\Delta^{(s)}_{2k}=\Delta^{(s)}_{3k}=0,\nonumber\\
&&\Delta^{(s)}_{5k}=B_3+2B_{cx}\gamma_{1k}+B_{cc}\gamma_{2k}.
\end{eqnarray}
Calculations show that for all realistic parameters of the model
there is no non-trivial solution. Consequently, in the SFM, when
the strong coupling of holes on oxygen ions with spin moments of
copper ions is correctly taken into account, the superconducting
phase with the $s$-wave symmetry of the order parameter is not
realized. This is the main difference between the theory of HTSC
developed here and the approaches based on the effective
single-band models of strongly correlated fermions on the square
lattice, in which along with the superconducting $d$-wave pairing
there is always a solution for the superconducting $s$-wave
pairing.

\section{CONCLUSION}\label{sec:conclusion}

The main results of this study can be formulated as follows.

(1) It has been shown that the neutralization for the negative
effect of the intersite Coulomb interaction of holes in the oxygen
subsystem on the Cooper instability in the $d$-wave channel occurs
as a result of two factors. The first factor is associated with
the analysis of the actual crystallographic structure of the
CuO$_2$ plane, according to which the Coulomb repulsion of
fermions in the oxygen sublattice is determined by the Fourier
transform of intersite interaction
$V_q=4V_1\cos(q_x/2)\cos(q_y/2)$. The second factor is associated
with the electron correlations leading to the emergence of strong
coupling between the localized spins of copper ions and holes at
the oxygen ions. As a result, the spin-polaron quasiparticles are
formed and move over the copper ion sublattice; in the ensemble of
these particles, the Cooper instability evolves. The Coulomb
repulsion between bare holes with the Fourier transform $V_q$ is
renormalized into the interaction between spin-polaron
quasiparticles so that the momentum dependence of this effective
interaction corresponds to the structure of the copper ion
sublattice. As a result, the situation takes place, when the
effective repulsion between the spin polarons falls out of the
equation for the superconducting order parameter with the $d$-wave
symmetry. At the same time, the contribution of such an effective
repulsion remains for the Cooper instability in the $s$-wave
channel.

(2) The solution of the system of self-consistent integral
equations for superconducting phases showed that in the
spin-fermion model only the phase with the $d$-wave symmetry of
the order parameter is realized, whereas solutions for the
$s$-wave pairing are not available for all the admissible levels
of doping. These results completely correspond to the experimental
data on cuprate superconductors. In this connection, we note that
within the $t-J$ model, the superconducting $s$-wave pairing is
realized, and the critical temperature corresponding to this phase
is much higher than $T_c$ for the $d$-wave pairing. Concerning the
differences that arise, it is appropriate to point out that in our
approach the spin subsystem of copper ions, separated from the
hole subsystem, plays an important role, whereas within the $t-J$
model the electron and spin degrees of freedom are due to the same
electrons.

(3) The effect of Coulomb repulsion $U_p$ for quasiparticles at
the same oxygen ion on the dependence of the superconducting
transition temperature for superconducting phase with the $d$-wave
symmetry of the order parameter on the doping level has been
analyzed. It is shown that taking $U_p$ into account leads to
decrease in the superconducting transition temperature, but this
temperature remains within the limits that are observed
experimentally.

It should also be noted that the different contributions of the
Coulomb interaction to the conditions of realization of the
superconducting phases with different symmetries of the order
parameter are manifested, for example, in the Kohn-Luttinger
theory of superconductivity~\cite{Kagan15}. In our case, the
separation factor plays a decisive role, when two types of oxygen
orbitals spatially separated from the spins of the copper ions are
taken into account.

\section{ACKNOWLEDGMENTS}

This work was supported by the Russian Foundation for Basic
Research (RFBR), the Government of Krasnoyarsk Region, the
Krasnoyarsk Region Science and Technology Support Fund (projects
nos. 16-42-240435, 16-42-243056 and 16-42-243057), and the Program
by SB RAS (project no. 356-2015-0406). The work of A.\,F.\,B. was
supported by the RFBR (project no. 16-02-00304). The work of
M.\,M.\,K. was supported by the Council for Grants of the
President of the Russian Federation (MK-1398.2017.2).

\section*{Appendix A}\label{sec:Appendix1}

Functions $S_{ij}^{(l)}(k,\omega)$ appearing in the expressions
for anomalous Green functions $F_{ij}(k,\omega)$ (\ref{Fij}) have
the form
\begin{eqnarray}
&&S^{(1)}_{11}(k,\omega)=Q_{3y}(k,-\omega)Q_{3y}(k,\omega),\nonumber\\
&&S^{(2)}_{11}(k,\omega)=S^{(1)}_{21}(k,\omega)= Q_{3}(k,-\omega)Q_{3y}(k,\omega),\nonumber\\
&&S^{(3)}_{11}(k,\omega)=S^{(1)}_{12}(k,\omega)= S^{(2)}_{11}(k,-\omega),\nonumber\\
&&S^{(4)}_{11}(k,\omega)=S^{(2)}_{12}(k,\omega)=S^{(3)}_{21}(k,\omega)=
S^{(1)}_{22}(k,\omega) \nonumber\\
&&\qquad\qquad=Q_{3}(k,-\omega)Q_{3}(k,\omega),\nonumber\\
&&S^{(5)}_{11}(k,\omega)=-Q_{y}(k,-\omega)Q_{y}(k,\omega),\nonumber\\
&&S^{(3)}_{12}(k,\omega)= Q_{3y}(k,-\omega)Q_{3x}(k,\omega),\nonumber\\
&&S^{(2)}_{21}(k,\omega)=S^{(3)}_{12}(k,-\omega),\nonumber\\
&&S^{(4)}_{12}(k,\omega)=S^{(3)}_{22}(k,\omega)= Q_{3}(k,-\omega)Q_{3x}(k,\omega),\nonumber\\
&&S^{(4)}_{21}(k,\omega)=S^{(2)}_{22}(k,\omega)=S^{(4)}_{12}(k,-\omega),\nonumber\\
&&S^{(5)}_{12}(k,\omega)=-Q_{y}(k,-\omega)Q_{x}(k,\omega),\nonumber\\
&&S^{(5)}_{21}(k,\omega)=S^{(5)}_{12}(k,-\omega),\nonumber\\
&&S^{(4)}_{22}(k,\omega)=Q_{3x}(k,-\omega)Q_{3x}(k,\omega),\nonumber\\
&&S^{(5)}_{22}(k,\omega)=-Q_{x}(k,-\omega)Q_{x}(k,\omega),\nonumber\\
&&S^{(1)}_{31}(k,\omega)=-K_kQ_{y}(k,-\omega)Q_{3y}(k,\omega),\nonumber\\
&&S^{(2)}_{31}(k,\omega)=-K_kQ_{x}(k,-\omega)Q_{3y}(k,\omega),\nonumber\\
&&S^{(3)}_{31}(k,\omega)=S^{(1)}_{32}(k,\omega)=-K_kQ_{y}(k,-\omega)Q_{3}(k,\omega),\nonumber\\
&&S^{(4)}_{31}(k,\omega)=S^{(2)}_{32}(k,\omega)=-K_kQ_{x}(k,-\omega)Q_{3}(k,\omega),\nonumber\\
&&S^{(5)}_{31}(k,\omega)=Q_{xy}(k,-\omega)Q_{y}(k,\omega),\nonumber\\
&&S^{(3)}_{32}(k,\omega)=-K_kQ_{y}(k,-\omega)Q_{3x}(k,\omega),\nonumber\\
&&S^{(4)}_{32}(k,\omega)=-K_kQ_{x}(k,-\omega)Q_{3x}(k,\omega),\nonumber\\
&&S^{(5)}_{32}(k,\omega)=Q_{xy}(k,-\omega)Q_{x}(k,\omega),\nonumber\\
&&S^{(1)}_{33}(k,\omega)=-K_k^2S^{(5)}_{11}(k,\omega),~S^{(2)}_{33}(k,\omega)=K_k^2S^{(5)}_{12}(k,-\omega),\nonumber\\
&&S^{(3)}_{33}(k,\omega)=S^{(2)}_{33}(k,-\omega),~S^{(4)}_{33}(k,\omega)=K_k^2S^{(5)}_{22}(k,\omega),\nonumber\\
&&S^{(5)}_{33}(k,\omega)=Q_{xy}(k,-\omega)Q_{xy}(k,\omega).
\end{eqnarray}
These expressions include the functions
\begin{eqnarray}
&&Q_{x(y)}(k,\omega)=(\omega-\xi_{x(y)})J_{y(x)}+t_kJ_{x(y)},\nonumber\\
&&Q_3(k,\omega)=(\omega-\xi_L)t_k+J_xJ_yK_k,\nonumber\\
&&Q_{3x(3y)}(k,\omega)=(\omega-\xi_L)(\omega-\xi_{x(y)})-J_{x(y)}^2K_k,\nonumber\\
&&Q_{xy}(k,\omega)=(\omega-\xi_x)(\omega-\xi_y)-t_k^2.
\end{eqnarray}

\section*{APPENDIX B}\label{sec:Appendix2}

Integrand functions $w_{ij}(k)$ defining matrix elements $W_{ij}$
in expression (\ref{def_W}) have the form
\begin{eqnarray}
&& w_{11}(k)=\zeta^{(5)}_k,~~~~~~~~~~
w_{12}(k)=\zeta^{(5)}_k\cos k_x,\nonumber\\
&& w_{13}(k)=\zeta^{(5)}_k\cos k_y,~~w_{14}(k)=\zeta^{(5)}_k\gamma_{2k},\nonumber\\
&& w_{15}(k)=\zeta^{(2)}_k\psi_k,~~~~~~~
w_{16}(k)=\zeta^{(3)}_k\psi_k,\nonumber\\ &&
w_{17}(k)=\zeta^{(1)}_k,~~~~~~~~~~
w_{18}(k)=\zeta^{(4)}_k,\nonumber
\end{eqnarray}
\begin{eqnarray}
&& w_{21}(k)=\zeta^{(5)}_{x,k},~~~~~~~~~~
w_{22}(k)=\zeta^{(5)}_{x,k}\cos k_x,\nonumber\\
&& w_{23}(k)=\zeta^{(5)}_{x,k}\cos k_y,~~
w_{24}(k)=\zeta^{(5)}_{x,k}\gamma_{2k},\nonumber\\
&& w_{25}(k)= \zeta^{(2)}_{x,k}\psi_k,~~~~~~~
w_{26}(k)=\zeta^{(3)}_{x,k}\psi_k,\nonumber\\
&& w_{27}(k)=
\zeta^{(1)}_{x,k},~~~~~~~~~~w_{28}(k)=\zeta^{(4)}_{x,k},\nonumber
\end{eqnarray}
\begin{eqnarray}
&& w_{31}(k)=\zeta^{(5)}_{y,k},~~~~~~~~~~
w_{32}(k)=\zeta^{(5)}_{y,k}\cos k_x,\nonumber\\
&& w_{33}(k)=\zeta^{(5)}_{y,k}\cos k_y,~~
w_{34}(k)=\zeta^{(5)}_{y,k}\gamma_{2k},\nonumber\\
&& w_{35}(k)=\zeta^{(2)}_{y,k}\psi_k,~~~~~~~
w_{36}(k)=\zeta^{(3)}_{y,k}\psi_k,\nonumber\\
&&
w_{37}(k)=\zeta^{(1)}_{y,k},~~~~~~~~~~w_{38}(k)=\zeta^{(4)}_{y,k},\nonumber
\end{eqnarray}
\begin{eqnarray}
&& w_{41}(k)= -V_1C_2\psi_k M^{(5)}_{ab}(k),~w_{42}(k)=w_{41}(k)\cos k_x,\nonumber\\
&& w_{43}(k)=w_{41}(k)\cos k_y,~~~~~~~w_{44}(k)=w_{41}(k)~\gamma_{2k},\nonumber\\
&& w_{45}(k)=-V_1C_2 M^{(2)}_{ab}(k)\psi^2_k,~w_{46}(k)=-V_1C_2 M^{(3)}_{ab}(k)\psi^2_k,\nonumber\\
&& w_{47}(k)=-V_1C_2
M^{(1)}_{ab}(k)\psi_k,~w_{48}(k)=-V_1C_2M^{(4)}_{ab}(k)\psi_k,\nonumber
\end{eqnarray}
\begin{eqnarray}
&& w_{51}(k)= -4V_1\psi_k M^{(5)}_{21}(k),~w_{52}(k)=w_{51}(k)\cos k_x,\nonumber\\
&& w_{53}(k)=w_{51}(k)\cos k_y,~~~~~w_{54}(k)=w_{51}(k)~\gamma_{2k},\nonumber\\
&& w_{55}(k)= -4V_1\psi_k^2 M^{(2)}_{21}(k),~w_{56}(k)= -4V_1\psi_k^2 M^{(3)}_{21}(k),\nonumber\\
&& w_{57}(k)= -4V_1\psi_k M^{(1)}_{21}(k),~w_{58}(k)= -4V_1\psi_k
M^{(4)}_{21}(k),\nonumber
\end{eqnarray}
\begin{eqnarray}
&& w_{61}(k)= -4V_1\psi_k M^{(5)}_{12}(k),~w_{62}(k)=w_{61}(k)\cos k_x,\nonumber\\
&& w_{63}(k)=w_{61}(k)\cos k_y,~~~~~w_{64}(k)=w_{61}(k)~\gamma_{2k},\nonumber\\
&& w_{65}(k)= -4V_1\psi_k^2 M^{(2)}_{12}(k),~w_{66}(k)= -4V_1\psi_k^2 M^{(3)}_{12}(k),\nonumber\\
&& w_{67}(k)= -4V_1\psi_k M^{(1)}_{12}(k),~w_{68}(k)= -4V_1\psi_k
M^{(4)}_{12}(k),\nonumber
\end{eqnarray}
\begin{eqnarray}
&& w_{71}(k)= -U_p M^{(5)}_{11}(k),~~~~w_{72}(k)=w_{71}(k)\cos k_x,\nonumber\\
&& w_{73}(k)=w_{71}(k)\cos k_y,~~~~w_{74}(k)=w_{71}(k)~\gamma_{2k},\nonumber\\
&& w_{75}(k)= -U_p M^{(2)}_{11}(k)\psi_k,~
w_{76}(k)= -U_p M^{(3)}_{11}(k)\psi_k,\nonumber\\
&& w_{77}(k)= -U_p M^{(1)}_{11}(k),~~~~w_{78}(k)= -U_p
M^{(4)}_{11}(k),\nonumber
\end{eqnarray}
\begin{eqnarray}
&& w_{81}(k)= -U_p M^{(5)}_{22}(k),~~~~w_{82}(k)=w_{81}(k)\cos k_x,\nonumber\\
&& w_{83}(k)=w_{81}(k)\cos k_y,~~~~w_{84}(k)=w_{81}(k)~\gamma_{2k},\nonumber\\
&& w_{85}(k)= -U_p M^{(2)}_{22}(k)\psi_k,~
w_{86}(k)= -U_p M^{(3)}_{22}(k)\psi_k,\nonumber\\
&& w_{87}(k)= -U_p M^{(1)}_{22}(k),~~~~w_{88}(k)= -U_p
M^{(4)}_{22}(k),\nonumber
\end{eqnarray}
where ($l=1,\dots,5$):
\begin{eqnarray}
& \zeta^{(l)}_k=2\xi_{t,x} M^{(l)}_{31}(k)+2\xi_{t,y}
M^{(l)}_{32}(k)-\frac34 V_1 \psi_{k}M^{(l)}_{ab}(k)
\nonumber\\
&-2J\gamma_{1k}M^{(l)}_{33}(k)+\left(\frac32 J
+8IC_1\right)M^{(l)}_{uu}(k) -\frac38 U_p
M^{(l)}_{aa}(k),\nonumber
\end{eqnarray}
\begin{eqnarray}
& \zeta^{(l)}_{x,k}=2I\cos k_x M^{(l)}_{33}(k)+V_1C_1\psi_k M^{(l)}_{ab}(k)\nonumber\\
& -2C_1(J+I\cos k_x)M^{(l)}_{uu}(k)+\frac12U_pC_1
M^{(l)}_{11}(k),\nonumber
\end{eqnarray}
\begin{eqnarray}
& \zeta^{(l)}_{y,k}= 2I\cos k_y M^{(l)}_{33}(k)
+V_1C_1\psi_kM^{(l)}_{ab}(k)
\nonumber\\
& -2C_1(J+I\cos k_y)M^{(l)}_{uu}(k)+\frac12U_pC_1
M^{(l)}_{22}(k),\nonumber
\end{eqnarray}
\begin{eqnarray}
& \xi_{t,x(y)}=\xi_{x(y)}s_{k,x(y)}+t_ks_{k,y(x)},\nonumber\\
& M^{(l)}_{uu}(k)=-s_{k,x}^2M^{(l)}_{11}(k)-s_{k,y}^2
M^{(l)}_{22}(k)-\psi_{k}M^{(l)}_{ab}(k),\nonumber\\
& M^{(l)}_{ab}(k)=M^{(l)}_{21}(k)+M^{(l)}_{12}(k),\nonumber\\
& M^{(l)}_{aa}(k)=M^{(l)}_{11}(k)+M^{(l)}_{22}(k).\nonumber
\end{eqnarray}

\end{document}